\documentstyle[sprocl,epsfig]{article}

\def\Journal#1#2#3#4{{#1} {\bf #2}, #3 (#4)}

\def\PRD{{\em Phys. Rev.} D}

\def\be{\begin{equation}}
\def\ee{\end{equation}}
\def\bea{\begin{eqnarray}}
\def\eea{\end{eqnarray}}

\begin{document}

\title{DIS STRUCTURE FUNCTIONS FROM THE SATURATION 
MODEL\footnote{Presented at the 8{th} International Workshop on Deep
Inelastic Scattering, Liverpool, 25-30 April 2000.}}

\author{K. GOLEC--BIERNAT\footnote{E-mail:golec@mail.desy.de}}

\address{II Institute of Theoretical Physics, Hamburg University,\\
Luruper Chaussee 149, D-22761 Hamburg, Germany\\ {\rm and}\\
Institute of Nuclear Physics, Radzikowskiego 152, \\
31-342 Krak\'ow, Poland}

\maketitle\abstracts{We present a description of 
inclusive and diffractive structure functions in DIS 
at small $x$, using a model based on high energy factorization. 
In this model the two processes have physical interpretation in
terms of the virtual photon wave function and the dipole cross section. 
We postulate the dipole cross section form in a way in
which unitarity is taken into account. A good description of data (including
DIS diffraction) is obtained after determining three parameters of the dipole
cross section from the fit to inclusive data only.}

DIS in the high energy limit $(x=Q^2/W^2\ll 1)$ can be given an
interpretation of a two step process. The virtual photon (emitted
by the incident electron) splits into a $q\bar{q}$ dipole which
subsequently interacts with the proton. The inclusive structure functions 
$F_2=F_T+F_L$ and $F_L$ are then given by 
\be
F_{T,L}(x,Q^2) = \frac{Q^2}{4\pi^2\alpha_{em}}
\int d^2{\bf{r}}\, dz\, |\Psi_{T,L}(r,z,Q^2)|^2\, \hat{\sigma}(x,r),
\ee
where $\Psi_{T,L}$ is the known {\bf wave function} for transverse (T) or
longitudinal (L) photon to split into a $q\bar{q}$ dipole, 
and $\hat{\sigma}$ is the {\bf dipole cross section} 
describing  the interaction of the dipole with the proton.
In addition, ${\bf{r}}$ is the transverse separation of the $q\bar{q}$ pair
and $z$ is the photon's momentum fraction carried by one of the quarks.
The computation of $\hat{\sigma}$ has been attempted within
perturbative QCD
assuming different types of the net colorless gluon exchange (e.g. DGLAP or
BFKL ladders, or multiple gluon interactions). Most of these attempts are
plagued, however,  by the problems with unitarity of finally computed cross
sections at small $x$. In our approach~\cite{ja}
we built in unitarity in the dipole
cross section by proposing the following phenomenological form
\be
\label{dcs}
\hat{\sigma}(x,r) = \sigma_0\,(1-\exp(-r^2/4 R_0^2(x)),
\ee
where $R_0(x)$, called the saturation radius, is given by
$R_0(x) = {1}/{Q_0}\,\cdot \left({x}/{x_0}\right)^{\lambda/2}$, with
$Q_0=1~\mbox{\rm GeV}$ and the parameters $\sigma_0, x_0$ and $\lambda$ 
fitted to all inclusive DIS data with $x<0.01$. At small $r$ Eq.~(\ref{dcs}) 
features color transparency and strong growth with $x$, 
$\hat\sigma\sim r^2\,x^{-\lambda}$.
For  large $r$
or $x\rightarrow 0$ the  constant value $\sigma_0$ is approached. The
transition to the saturated form is governed by $R_0(x)$.

The presented model successfully describes $F_2$ at small $x$, 
see Fig.~1(left), with a particular emphasis on the transition between
small and large $Q^2$. In addition, the found dipole cross section was 
successfully applied to the  description of DIS diffraction~\cite{ja}, see
Fig.~1(right). The constant ratio between the diffractive and inclusive cross
sections also finds an explanation in this model. For related approaches see
\cite{shaw,mcd} and  for a possible theoretical explanation see \cite{kov}.

\begin{figure}[t]
\begin{center}
\epsfig{figure=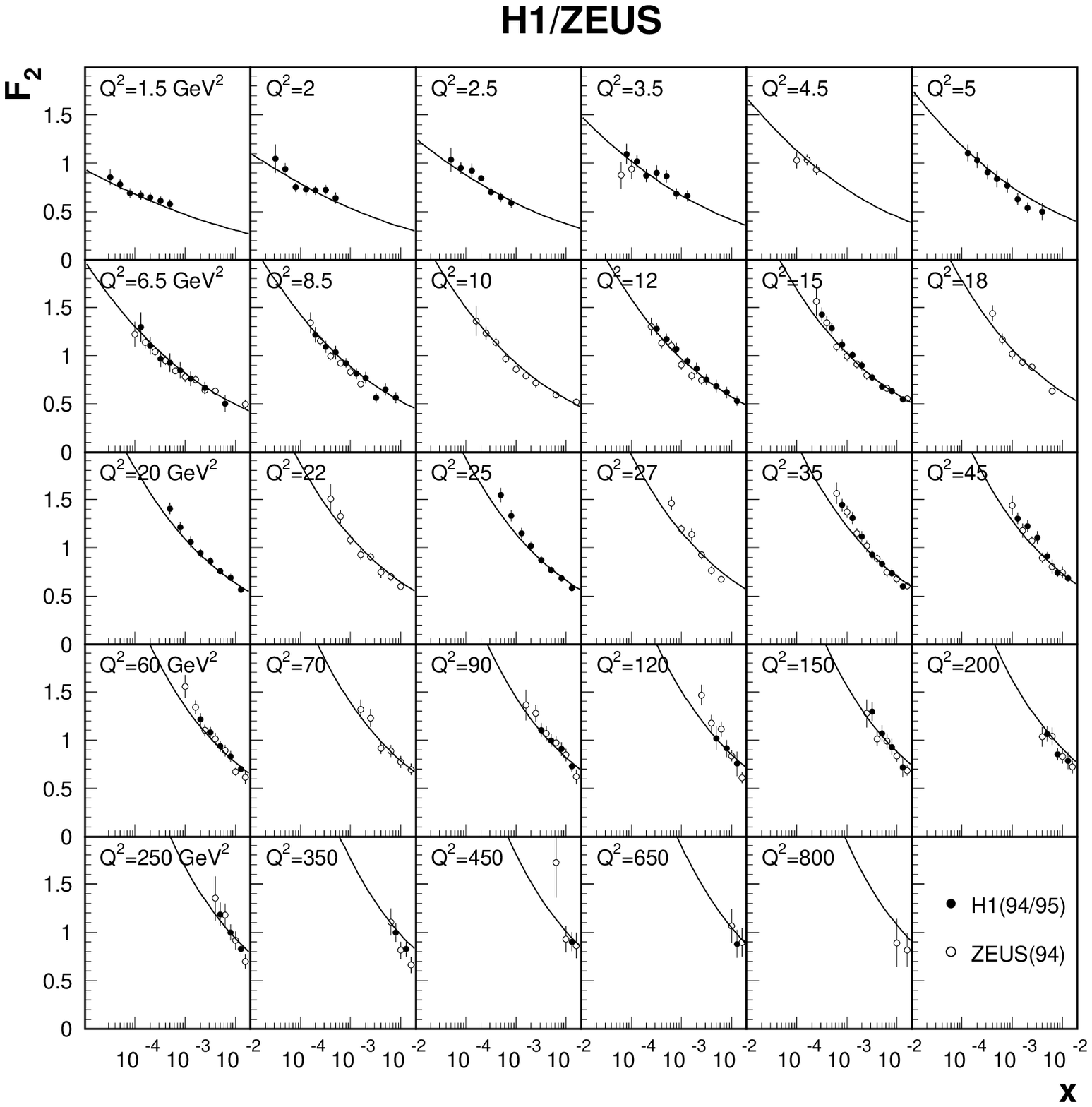,width=5.5cm}
\epsfig{figure=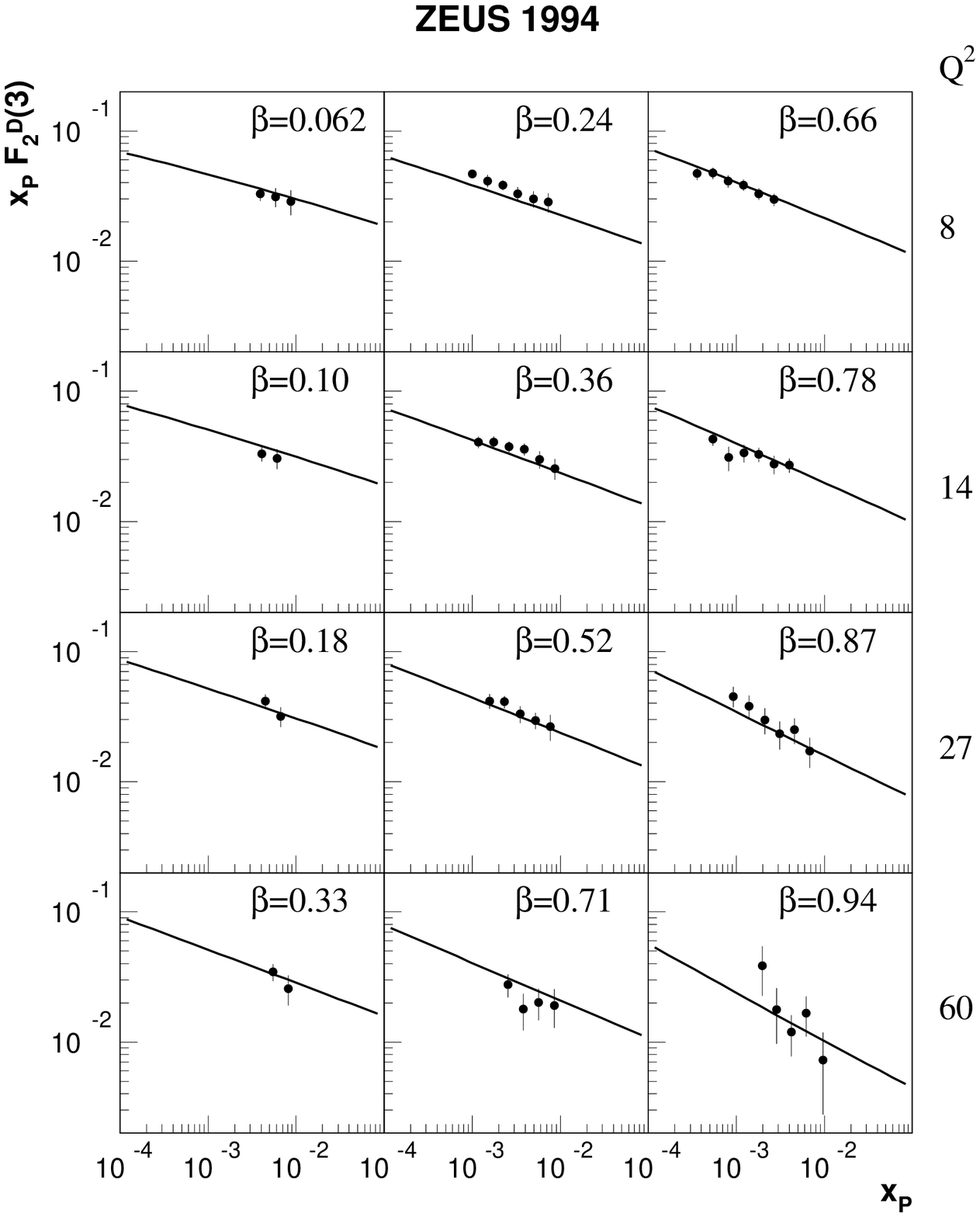,width=5.5cm}
\end{center}
\vspace*{-0.5cm}
\caption{$F_2$ and  diffractive $F_2^{D(3)}$
structure functions in the saturation model.}
\end{figure}

\vspace*{-0.4cm}
\section*{Acknowledgments}
This work was done in collaboration with Mark W\"usthoff.
Supported by {\it Deutsche Forschungsgemeinschaft}
and Polish KBN grant no. 2 P03B 089 13. 


\end{document}